\documentclass[12pt,preprint2]{aastex61}
\usepackage{amsmath,amssymb}
\usepackage{footnote}
\usepackage{float}
\usepackage{color}
\usepackage[utf8]{inputenc}
\usepackage{graphicx}

\usepackage{aas_macros}
\usepackage{hyperref} 
\hypersetup{colorlinks,citecolor=blue,linkcolor=blue,urlcolor=blue}

\hypersetup{draft}
\shorttitle{Coarse-grained Complexity}
\shortauthors{Segal et al.}
\begin{document}
\title{Identifying complex sources in large astronomical data using a coarse-grained complexity measure}
\author{Gary Segal} 
\affiliation{School of Mathematics and Physics, University of Queensland, St Lucia, Brisbane, QLD 4072, Australia}%
\affiliation{CSIRO Astronomy and Space Science, PO Box 76, Epping, 1710, NSW, Australia}
\email{g.segal@uq.edu.au}
\author{David Parkinson}
\affiliation{School of Mathematics and Physics, University of Queensland, St Lucia, Brisbane, QLD 4072, Australia}%
\affiliation{Korea Astronomy and Space Science Institute, Daejeon 34055, Korea}
\email{davidparkinson@kasi.re.kr}
\author{Ray P Norris}
\affiliation{Western Sydney University, Locked Bag 1797, Penrith South, 1797, NSW, Australia}%
\affiliation{CSIRO Astronomy and Space Science, PO Box 76, Epping, 1710, NSW, Australia}
\email{Ray.Norris@csiro.au}
\author{Jesse Swan}
\affiliation{School of Natural Sciences, University of Tasmania, Private Bag 37, Hobart 7001, Australia}%
\affiliation{CSIRO Astronomy and Space Science, PO Box 76, Epping, 1710, NSW, Australia}

\begin{abstract} 
The volume of data that will be produced by the next generation of astrophysical instruments represents a significant opportunity for making unplanned and unexpected discoveries. Conversely, finding unexpected objects or phenomena within such large volumes of data presents a challenge that may best be solved using computational and statistical approaches. We present the application of a coarse-grained complexity measure for identifying interesting observations in large astronomical datasets. This measure, which has been termed apparent complexity, has been shown to model human intuition and perceptions of complexity. Apparent complexity is computationally efficient to derive and can be used to segment and identify interesting observations in very large datasets based on their morphological complexity. We show using data from the Australia Telescope Large Area Survey (ATLAS) that the apparent complexity can be combined with clustering methods to provide an automated process for distinguishing between images of galaxies which have been classified as having simple and complex morphologies. The approach generalises well when applied to new data after being calibrated on a smaller dataset, where it performs better than tested classification methods using pixel data. This generalisability positions apparent complexity as a suitable machine learning feature for identifying complex observations with unanticipated  features. 

\end{abstract}


\section{Introduction}

The Universe is very large, and we have only just begun to scratch the surface in terms of identifying the different objects and events that it contains. Each generation of instrumentation and infrastructure has expanded our knowledge and sample size significantly, often resulting in unexpected scientific results. For example, of the 10 greatest discoveries by the Hubble Space Telescope, only one was listed in its key science goals \citep{norris2017}. The next generation of astrophysical instruments will be collecting petabytes of data per day. This represents a significant opportunity for making unplanned discoveries due to the large volume of observational data that will be made available, but is beyond the limit for the amount of information that can be examined directly by the human astronomical community on any reasonable timescale. 

One good example of this expansion in surveying and collection capability is in the area of extragalactic radio astrophysics. We currently know of about 2.5 million extragalactic radio sources, but future surveys will increase this number by several orders of magnitude. The Australian Square Kilometre Array Pathfinder (ASKAP)  and the Evolutionary Map of the Universe (EMU) survey is predicted to increase this number to about 70 million \citep{norris_2017a} in the continuum. The Square Kilometre Array (SKA) may increase this number into the billions \citep{SKA2015Jarvis}.

Progress that has been made with supervised machine learning approaches such as Convolution Neural Networks have demonstrated the potential effectiveness of these approaches for identifying and classifying observations in astronomical surveys based on their features \citep{Thorat,karpenka,kimbailerjones,desspcc,Dieleman,Huertas-Company,Charnock, alger2018}. Supervised approaches may become highly effective at identifying observations that have been previously considered interesting, utilising features associated with the interesting observations used for training and testing, but overlooking new observations whose features have little in common with past interesting observations. Such an approach can be said to suffer from an expectation bias discussed by \citet{norris2017} and \citet{Robinson87}. Unsupervised learning methods, or the use of selected features that have been demonstrated to generalise well, may therefore be considered preferable candidates for segmenting new observations and detecting the unexpected.

A good example of an unsupervised learning approach for finding outliers inside the ensemble was the use of random forests by \citet{Baron}. Here they have utilised the correlation structure in feature space to identify interesting observations using learned features based on the correlation structure of galaxy spectra. They used random forests to identify interesting features in galaxy spectra, by learning the difference between real and synthetic observations, where the correlation structure was removed from the feature space in the synthetic case. Outliers were then identified based on a measure of similarity between objects, where the authors counted how often every pair of real objects were classified as real in the same leaf of a given tree. This approach demonstrates the use of learned interesting features, but learning not based on previous observations, to perform subsequent outlier detection. A potential drawback of this approach, as with many other unsupervised outlier detection methods, is the computational burden of having to compute pairwise comparisons. The use of selected features, that generalise well, present a computationally efficient option when applying machine learning approaches to segment the large volumes of data being processed in the early stages of the image analysis pipeline.

This paper presents the application of a coarse-grained complexity measure for identifying observations with complex and interesting morphologies in large astronomical datasets. \citet{Carroll} have demonstrated the effectiveness of a coarse-grained complexity measure, termed apparent complexity, at capturing human intuition and perceptions of complexity. This measure provides a quantitative description of a notion of complexity informally proposed by \citet{GellMann94} as a phenomena that first increases and then decreases with the rising entropy of a closed system. A potential drawback of apparent complexity as a formal description of complexity is that it relies on assumptions regarding human perceptions. Conversely, this very connection to human perceptions suggests that this measure should be effective at identifying complex observations that are likely to be of interest to a human observer. This paper shows that by learning an appropriate smoothing function, the apparent complexity can be used to partition a sample based on the morphological complexity and interestingness of observations.

We present the application of apparent complexity as a machine learning feature, which can be used by itself or together with the signal to noise ratio to effectively identify radio galaxies with complex morphologies in the presence of noise. We show that the measure generalises well when applied to new and larger datasets with a more expansive range of varied morphologies. We also compare approaches using the measure to more traditional classification approaches, logistic regression and support vector machines (SVM) applied to pixel data, and show that the use of the complexity measure generalises comparatively better when applied to new and larger data after being calibrated on a smaller dataset.

Apparent complexity is fast and computationally efficient to compute, as the measure only requires applying a smoothing function and compression algorithm, both of which can be implemented at worst case linear time complexity:\begin{equation}T(n) = O(n)\end{equation} Implementations of constant time $O(1)$ and linear time $O(n)$ median filters are detailed in \citet{Perreault}. Being able to leverage fast and efficient tools that generalise well is likely to be desirable when identifying interesting and unexpected observations in very large scientific datasets such as those that will be produced by ASKAP and the SKA.

We envisage a potential application for the apparent complexity measure in the early stages of the image analysis pipeline to segment very large datasets and identify smaller samples of complex radio sources that are appropriate for additional analysis requiring more computationally intensive methods or manual inspection. This is likely to include cross matching with observations at other frequencies to determine if and how the host galaxy is coincident with radio components. It may also serve in the early identification of new and unexpected observations. 

Using data from the Australia Telescope Large Area Survey (ATLAS, \citet{Norris06}) we show that apparent complexity can be used to distinguish between images of galaxies which have been classified as having simple and complex morphologies. We also show that the approach generalises well when applied to new data.

The paper is structured as follows: section \ref{sec:theory} frames the theory in terms of Kolmogorov complexity and discusses the theoretical merits of apparent complexity as an attractive candidate for identifying interesting astronomical observations in large datasets. Section \ref{sec:results} outlines empirical methods and results that show apparent complexity can be used to distinguish between images of galaxies with simple and complex morphologies. It will also show that a smoothing function calibrated on a small labelled sample with few interesting observations is able to generalise well when applied to a much larger sample containing a larger collection of complex morphologies. Testing on data containing a larger sample of simple and complex observations than that used for training, we show that clustering using of the complexity measure generalises better than traditional classification methods using pixel data. Finally in section \ref{sec:conclusions} we summarise our conclusions.

\section{Theory}
\label{sec:theory}
In this paper we measure the complexity $C$ of some observation $x$, subject to a function $f$ that extracts only the non-incidental information from measurements. As such, the measured complexity will depend on the function $f$, that can be calibrated to align with the interests and perceptions of the scientific observer, and so should be considered as $C(f(x))$. It is envisioned that the association between the computed measure $C(f(x))$ and the morphological complexity of observations will generalise well.

\subsection{Apparent Complexity}
\label{sec:apparentcomplex}

Apparent complexity has been defined by \citet{Carroll} as the entropy H of an object $x$ after applying a smoothing function f, $H (f (x))$. The Shannon entropy of a probability distribution P can be defined as the expected number of random bits that are required to produce a sample from that distribution:
\begin{equation}H(P) = -\sum_{x \in X} P(x)\log P(x) \,. \end{equation}

By Shannon's Noiseless Coding Theorem the minimum average description length $L$ of a sample is close to the Shannon entropy:
\begin{equation}H(P) \leq L \leq H(P) + 1 \,. \end{equation}
The Kolmogorov complexity $K (f (x))$  can be used as a proxy for the entropy of the smoothed function $H (f (x))$, as proposed by \citet{Carroll}. The seeming analogy between the concept of entropy and program size has been previously recognised \citep{Chaitin}. The Kolmogorov complexity, or prefix complexity, of $x$ is the length of the shortest binary program $l(p)$, for the reference universal prefix Turing machine $U$, that outputs $x$; it is denoted as $K(x)$:
\begin{equation}K(x) = {\rm min}_{p}\{l(p):U(p)=x\} \,. \end{equation}

A thorough treatment is provided by \citet{Li}. The Kolmogorov complexity has the advantage of being well-defined for a particular description of a system such as an image of a galaxy. This is not the case for the Shannon entropy which is defined in terms of the possible states of the system. While the Kolmogorov complexity is uncomputable, its upper bound can be reasonably approximated by the compressed file size $C (f (x))$ using a standard compression program \citep{Carroll}, such as \texttt{gzip}. 

The issue with using the approximated Kolmogorov complexity directly as measure of complexity is that it is maximized by random information. Intuitively a complexity measure should provide low values for random data that does not contain structure that is of interest to the observer \citep{Zenil}. \citet{Carroll} have shown that the apparent complexity measure is able to achieve this by applying a smoothing function $f$ to the input $x$. 

While the Kolmogorov complexity of a random sequence is large, the apparent complexity of the same sequence becomes small with smoothing, as fluctuations are removed where the average or median information content becomes homogeneous at the coarse-grained resolution. Accordingly, we define the apparent complexity as the compressed description of regularities and structure after discarding all that is incidental. The apparent complexity will be small for both simple and random sequences.

The objective of applying the smoothing function $f$ when deriving $C(f(x))$ is to remove incidental or random information, such as instrumental noise, that is incomprehensible to the observer even though it may have a physical basis. Comprehensibility here is defined with respect to the observer of information, in this case scientists with specific interests. Comprehensible information has a structure within feature space, which in the case of images refers to the spatial distribution of bits of information across available channels. The apparent complexity measure can be calibrated to align with expert distinctions between meaningful structure and noise by adjusting the measurement resolution through a  smoothing function so that complexity values correctly partition expertly labelled data.

The apparent complexity measure of an image does not rely on the presence of any particular structures or structural elements and is invariant to rigid motions of the plane. The measure makes only explicit assumptions regarding the choice of coarse-graining level and the scale of the image. Previous data is therefore used only to calibrate the coarse-graining level (i.e. the appropriate measurement resolution).

Apparent complexity runs into obstacles as a well-defined measure of complexity. Firstly, the uncomputability of the Kolmogorov complexity prohibits the concept from being defined in terms of an optimal compression. It has been proven by \citet{Chaitin1} that there can be no procedure for finding all theorems that would allow for further compression. Furthermore the problem of distinguishing between meaningful structure and incidental information, especially in finite data, may fail to be well-defined. Different smoothing functions and different coarse-graining levels will retain different distinct regularities in the data.

These theoretical challenges in objectively defining the apparent complexity can be circumvented when the approach is applied to the segmentation of observations by complexity. Here the apparent complexity can be calibrated to coincide with notions of complexity adopted by the observer.

\section{Analysis of radio continuum data}
\label{sec:results}

\begin{figure}[h] 
\includegraphics[width=8cm]{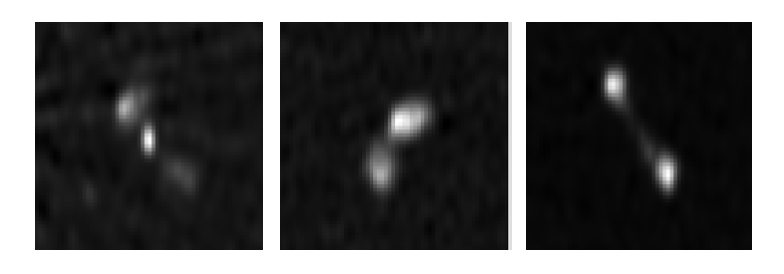}
\caption{\label{fig:complex} Complex radio sources with multiple components}
\end{figure}

\begin{figure}[h] 
\includegraphics[width=8cm]{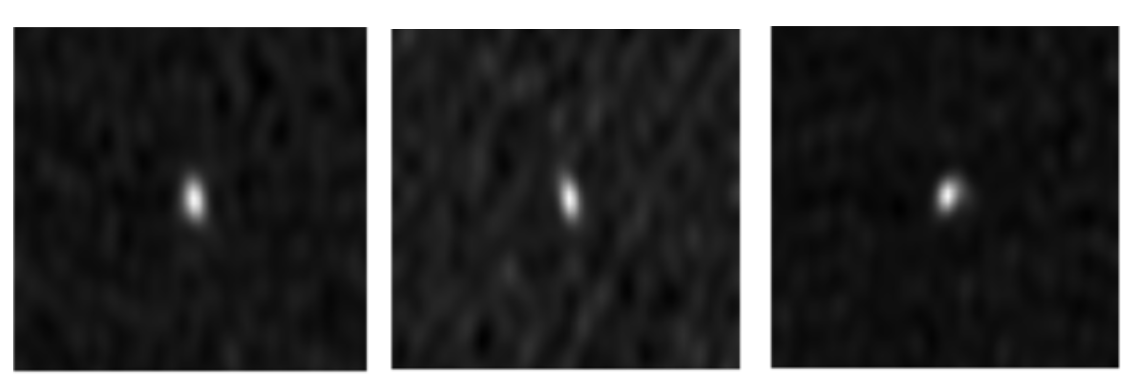}
\caption{\label{fig:simple} Simple unresolved radio sources}
\end{figure}

Segmentation based on apparent complexity can be used to identify complex images or complex regions within an image.
We demonstrate this approach using radio continuum images from the Australia Telescope Large Area Survey (ATLAS) survey, to distinguish between simple and complex radio sources. Here we define ``simple" sources as single unresolved components, and ``complex" sources as anything else, including bent-tail galaxies and extended radio sources (e.g. Fanaroff-Riley I, Fanaroff-Riley II) containing bright radio components in combination with diffuse plume-like jets. Figure \ref{fig:complex} provides examples of complex radio sources and figure \ref{fig:simple} provides examples of simple sources. 

It would be expected that two simple radio sources (unresolved sources), representing spatially separated galaxies that are randomly associated, will contain a significant amount of shared information in their morphologies, due to the similarity of the basic components. Conversely a true complex source is likely to contain components of a differing nature, such as radio lobes of differing luminosity and jets and plumes with luminosity gradients. By the unique decompression property distinct components will require additional bits of information in the combined compressed description \citep{Cilibrasi}. Accordingly the apparent complexity of a complex radio source should be larger than the apparent complexity of two simple sources in proximity.

The measure of apparent complexity can be used as a proxy for how interesting a radio image of a galaxy is, on the basis that radio sources with a larger apparent complexity contain a larger number of distinct and meaningful components that are likely to be of interest to an observer. We first demonstrate this intuition by distinguishing between artificially generated doubles and true complex sources, before using the measure to segment the ATLAS data with clustering methods.

\subsection{Method}
\label{sec:method}

We approximate the apparent complexity measure by applying a median filter $f$ with a window size calibrated to 10 pixels to an image $x$, and then calculating the \texttt{gzip} \citep{gzip} file size $C (f (x))$ as an upper bound on the Kolmogorov complexity $K (f (x))$. We apply this approach to the 256 by 256 pixel radio continuum images as follows:

\begin{enumerate}
  \item Load a centred image as a 256 by 256 matrix of 8 bit channel pixel intensity values
  \item Crop from the centre of the image to create a 64 by 64 matrix
  \item Filter the matrix using a percentile based threshold (P90) for pixel intensity values
  \item Apply a median filter using a learned window size ($h$ = 10) to produce a smoothed 64 x 64 matrix  (to remove random information and retain structural information)
  \item Compress the smoothed array using \texttt{gzip}
  \item Measure the compressed image size to estimate an upper bound of the Kolmogorov complexity
\end{enumerate}

We adopt a median filter with window size calibrated at $h$ = 10 pixels, using a square structuring element with data extended at image borders through reflection. The choice of window size is the only free parameter in this method, and is learned from some small training set, as described in section \ref{sec:calibration}. 

The median filter was selected because it completely removes noise and incidental values in regions predominately without flux measurement and retains the strength of signals in regions dominated by actual flux measurements. It is also better at preserving edges (than, for example, a Gaussian filter) given the expected salt-and-pepper noise.

Images were filtered based on a percentile based threshold for pixel intensity values set at the 90th percentile. The 90th percentile (P90) threshold was chosen through visual inspection of the images based on the removal of noise and imaging artifacts concentrated at lower flux values.  Appendix B shows the the difference in average apparent complexity between simple and complex ATLAS DR1 images across a range of measurement resolutions at different percentile based thresholds, namely, the P40, P50, P60, P70, P80 and P90. At thresholds lower below P60 the images become noise dominated and effective separation between simple and complex images is no longer achieved through smoothing. These results support the appropriateness of a large threshold, such as the P90, which provides both reasonable estimation and clearer separation of the average apparent complexity for simple and complex sources. This parameter has not been finely tuned however, and alternative threshold values my provide superior performance. Furthermore, this value was held fixed during calibration and testing so that the measurement resolution was the only free parameter. 

Images were cropped to produce a 64x64 matrix. The matrix size was selected to retain all information (measurements) associated with the source of interest while reducing the image size to facilitate faster and more efficient processing.

The gzip compression software was selected to provide lossless compression post-smoothing. This allows information loss to be calibrated through the smoothing process. The gzip software was used due to its speed, broad usage and its adoption in prior work by \citet{Carroll}.

\subsection{Survey sample}

ATLAS data consists of deep radio continuum imaging of the the Chandra Deep Field South (CDFS) and the European Large Area ISO Survey (ELAIS). The data is described in  \citet{Norris06} \& \citet{Middelberg08} (DR1), and \citet{Franzen15} (DR3).   

Table \ref{tab:tri} shows the number of identified radio sources within each field and data release. The table also provides a breakdown between sources that have been classified by human inspection from \citet{Norris06} and Norris et al (direct communication) as having simple and complex morphology. 

\begin{table}
\begin{tabular}{ l  c  c }
\hline \hline
DR3 (n=4825)& Complex obs & Simple obs \\
\hline 
ELAIS & 72 & 1892\\
CDFS & 97 & 2764 \\ 
\hline \hline
DR1 (n=708)& Complex obs & Simple obs \\
\hline 
CDFS & 34 & 674\\
\hline
\end{tabular}
\caption{ATLAS DR1 and DR3 samples}
\label{tab:tri}
\end{table}

The sources were provided as 256x256 pixel Portable Network Graphics (PNG) files. The images provided were pre-processed as detailed in \citet{Norris06,Franzen15}.

Labels were provided for ATLAS data release 1 (DR1) files identifying which sources had been classified as simple and complex. The data was then used to learn the smoothing function window size using images of galaxies that were manually labeled as having complex and simple morphologies. The window size of the smoothing function was chosen as to maximise the difference between the average apparent complexity of observations labelled complex and simple in the ATLAS DR1 sample.  

As shown in table \ref{tab:tri}, the ATLAS DR3 data provides a much larger sample containing more complex sources. The analysis for data release 3 (ATLAS DR3), was conducted `blind', where labels were not provided with the source files. The success of the approach could therefore be judged independently using the DR3 data.

\subsection{Smoothing function calibration}
\label{sec:calibration}

The appropriate window size for a smoothing function can be learned from a training set by maximising the difference between the apparent complexity of observations expertly labelled as complex and simple. As suggested by \citet{Carroll} there may also be natural choices for selecting the smoothing function suggested by our physical ability to actually observe systems and our knowledge of the systems properties.

\begin{figure}[h]
\includegraphics[width=8cm]{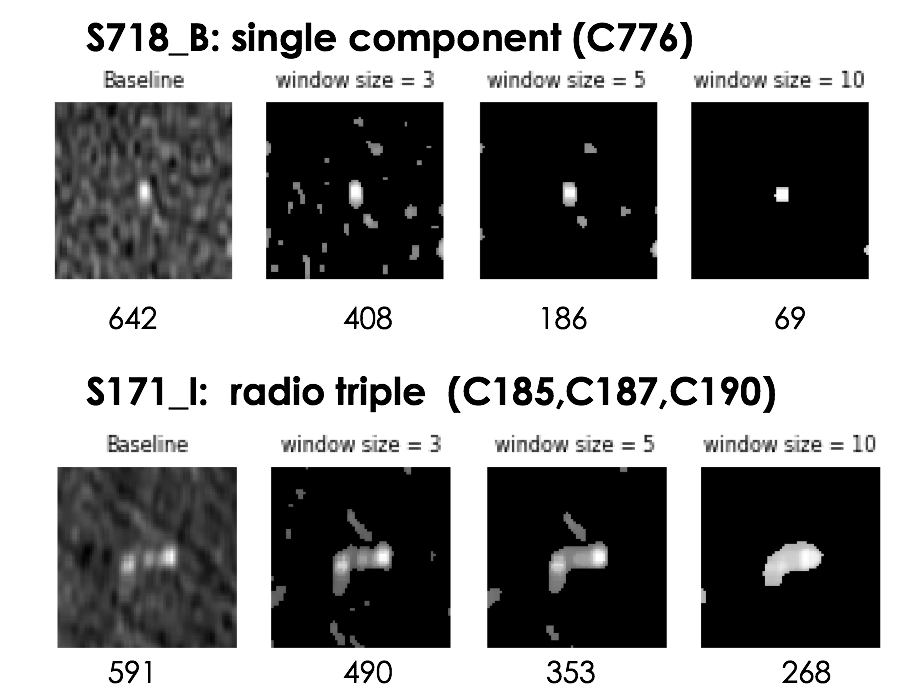}
\caption{\label{fig:windowsizes}ATLAS DR1 sources with progressive smoothing function window sizes and complexity scores in bytes}
\end{figure}

Applying an appropriate smoothing function to the radio images appears to remove random information and retain information contained within a structure comprised of distinct regions with different mean or median pixel values. The isolation of structure in a complex source is shown through the progressive application of coarser median filters in figure \ref{fig:windowsizes}. Images containing more random information show an initial rapid reduction in apparent complexity. 

\begin{figure}[h]
\includegraphics[width=8cm]{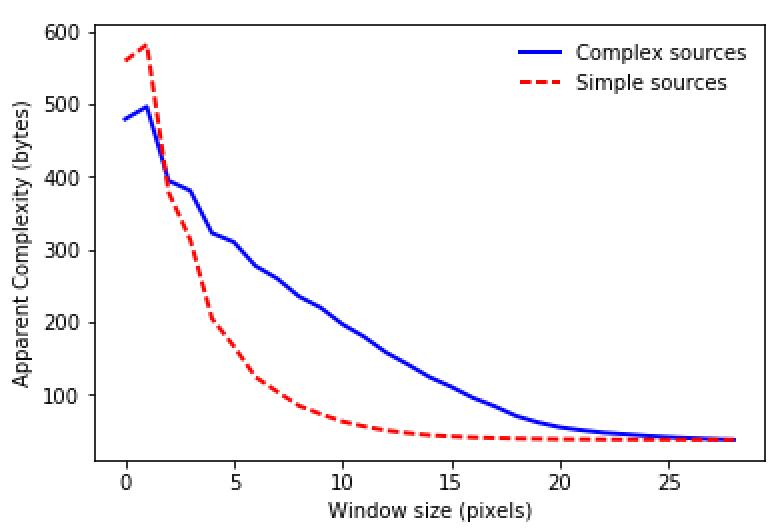}
\caption{\label{fig:smoothingscale}A comparison of the average apparent complexity between simple and complex ATLAS DR1 sources across changes in the size of the smoothing function window.}
\end{figure}

Figure  \ref{fig:smoothingscale}  compares the average apparent complexity of complex and simple ATLAS DR1 images across changes in the smoothing function window size. As the smoothing function window size increases, the apparent complexity of the simple and complex sources decreases, since information is being removed, but importantly they decrease at different rates. The near exponential shape of the curve representing simple sources suggests that the measured complexity of these images consists of random information at the baseline pre-processing level. Conversely, the changing apparent complexity of the images labelled as complex, that reverts to a closer to linear rate of decrease, suggests a greater content of coarser, more comprehensible, information. Where the apparent complexity curves of both complex and simple sources become flatter and converge, the smoothing function window size is large enough to remove both random and comprehensible content.

This analysis shows that there is an appropriate smoothing function window size that provides a clear separation, on average, between images that have been expertly classified as complex and simple sources in the ATLAS DR1 data. The separation between the average apparent complexity values of simple and complex sources becomes clear between a window size of 5 to 25 pixels, with the largest separation achieved toward the centre of this range. We fixed our choice to a diameter of 10 pixels to use when applying this function to the ATLAS DR3 data. The separation disappears at larger coarse-graining levels where the smoothing filter blacks out the image, removing all content.  At null and low smoothing levels the images represent the baseline pre-processing level of the ATLAS images. 

It is interesting to note that at the baseline pre-processing level, the average apparent complexity of images representing simple sources exceeds the average apparent complexity of images representing complex sources. The rapid reduction in complexity of these images with increased smoothing suggest that this higher initial apparent complexity is likely to be attributable to a higher content of random information. This is in contrast to the truly complex images that retain a greater percentage of the measured complexity at equivalent smoothing levels.

\subsection{Experimental Results}

\begin{figure}[h]
\includegraphics[width=8cm]{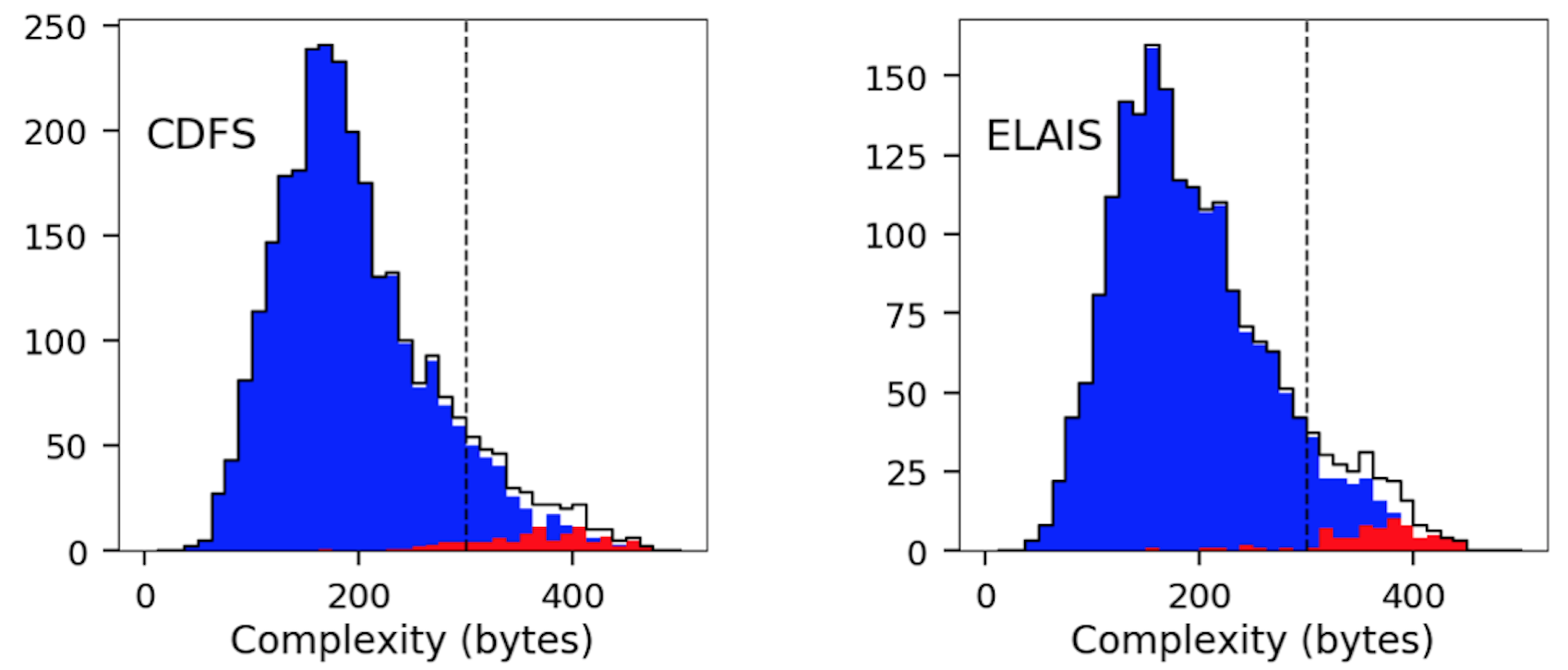}
\caption{\label{fig:complexity_hist}Distributions of the apparent complexity of the total sample (black solid line), subdivided into  simple (blue) and complex (red) radio sources (human classified) in the CDFS (\textit{left}) and ELAIS (\textit{right}) fields from ATLAS DR3. The black dashed vertical line gives the 300 byte boundary we assume in order to partition the two populations by apparent complexity. }
\end{figure}

We computed the apparent complexity of the ATLAS DR3 images, using the method described in section \ref{sec:method}. The resulting distribution of values is shown in figure \ref{fig:complexity_hist}, showing that the apparent complexity values for expertly classified simple and complex sources slightly overlap but with complex sources concentrated in the heavier right tails of the distributions.  

Without reference to the expertly classified labels for the images, we adopted a threshold of the approximate 90th percentile of the apparent complexity distribution (of the combined sample) segmenting the heavy tails that we interpreted as being influenced by a second overlapping distributions assumed to represent complex sources. In this way we assume a segmentation boundary at an apparent complexity threshold of 300 bytes, equivalent to selecting approximately the top 10\% of complexity values. 

The performance of this binary segmentation with respect to the true nature (expert classification) are shown in table \ref{tab:confusionmatrixcomponly}. These results show that the apparent complexity measure, at the selected partition boundary, can be used to correctly identify 86\% (i.e. a recall of 0.86) of the interesting observations from the combined DR3 samples with a 91\% reduction in the non-interesting data volume following classification.

Applying clustering methods to the complexity measurements allows us to implement an automated process for detecting interesting and unexpected observations. We apply clustering methods to the complexity measurements to avoid reliance on a large training set that would be needed to represent the feature space and class boundaries. To demonstrate the generality of the apparent complexity measure we calibrate the smoothing function on a small sample (ATLAS DR1, n=708) and test it using a much larger sample (ATLAS DR3, n=4825).

These considerations reflect the use case for an approach that is intended to identify complex observations, with often unexpected features, in new data that is extensively larger and more varied than any previous datasets, such as, the new data that will be produced by next generation instrumentation such as ASKAP and the SKA.

To demonstrate an automated partitioning process we performed cluster analysis by fitting a two component Gaussian mixture model (GMM) using the expectation maximisation (EM) algorithm. Binary segmentation using this approach is shown in table \ref{tab:confusionmatrixcompGMM}. Results based on this approach correctly identify 82\% (i.e. a recall of 82) of the interesting observations from the combined DR3 samples with a 94\% reduction in the non-interesting data volume following classification.

\begin{table}
\begin{tabular}{ l  c  c }
\hline \hline
CDFS (n=2861)& Complex obs & Simple obs\\
\hline 
Prediction: Complex& 81 & 244\\
Prediction: Simple& 16 & 2520 \\ 
\hline \hline
ELAIS (n=1964)& Complex obs & Simple obs\\
\hline 
Prediction: Complex& 65 & 167\\
Prediction: Simple& 7 & 1725\\  
\hline
\end{tabular}
\caption{\label{tab:confusionmatrixcomponly}Confusion matrix (P90 complexity cut)}

\end{table}

\begin{table}
\begin{tabular}{ l  c  c }
\hline \hline
CDFS (n=2861)& Complex obs & Simple obs\\
\hline 
Prediction: Complex& 73 & 141\\
Prediction: Simple& 24 & 2623 \\ 
\hline \hline
ELAIS (n=1964)& Complex obs & Simple obs\\
\hline 
Prediction: Complex& 65 & 149\\
Prediction: Simple& 7 & 1743\\  
\hline
\end{tabular}
\caption{\label{tab:confusionmatrixcompGMM}Confusion matrix (Complexity partition based on GMM )}

\end{table}

Type II errors, representing the incorrect classification of complex sources as simple, may be due to the removal of meaningful information by the smoothing function or potentially by the sparse representation of complex features, discernible to a human observer, but having little impact on the information content of the image. Classification errors may also be attributable to the allocation of the partition boundary. 

Type I errors, representing the incorrect identification of simple sources as complex, may be due to the presence of non-random information deemed by a human observer to be incidental and not contributing to the complexity of the source itself. An example could be a telescope imaging artifact containing structure, such as a point spread function originating from a brighter source. Examples of telescope imaging artifacts include the diagonal lines shown at baseline in figure \ref{fig:windowsizes}. 

Alternatively, type I errors may be explained by the retention of random information not removed by the smoothing function. Figure \ref{fig:smoothingscale} shows that there is a large amount of random information in the simple sources at baseline, and this suggests there is a risk that in some images random information will produce incidental structure that may not be removed through smoothing. Where random information is retained after smoothing, segmentation is likely to be improved by incorporating thresholds in both the apparent complexity and the signal-to-noise ratio (SNR), as random information is likely to be distributed more uniformly across the available channel intensity values. The SNR can be calculated as the reciprocal of the coefficient of variation for the channel intensity values.

\begin{figure}[h]
\includegraphics[width=8cm]{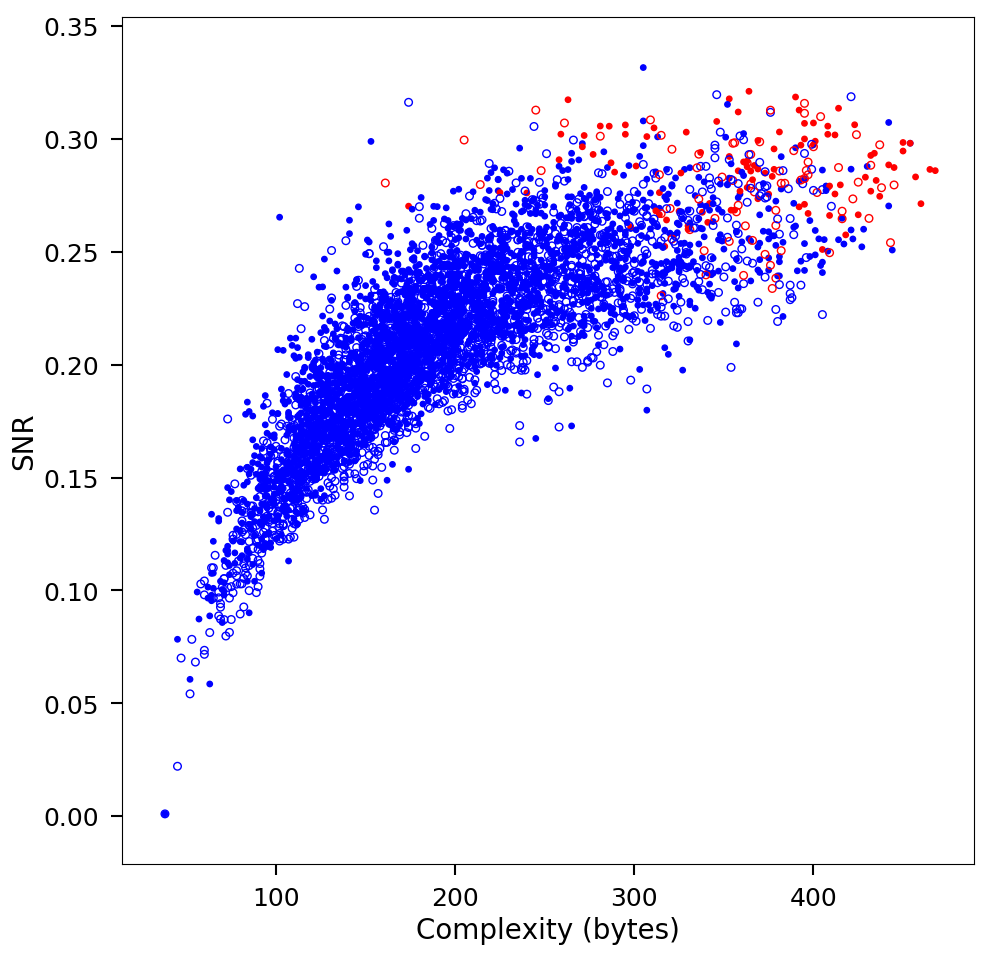}\\
\caption{\label{fig:complexity_snr_results}Scatter plot demonstrating the effectiveness of apparent complexity and SNR to partition simple (blue) and complex (red) radio sources in the CDFS (filled circle) and ELAIS (empty circle) fields from ATLAS DR3.}
\end{figure}

Figure \ref{fig:complexity_snr_results} shows both the apparent complexity and the SNR for the the ATLAS DR3 samples, and the relationship between them, where the samples have been partitioned into simple and complex by human inspection. In this figure the true complex radio sources are clustered at larger apparent complexity and SNR values, towards the top right edge of the scatter plot. This figure shows the extra information provided by the SNR in segmenting simple and complex sources.  

We automated the segmentation process by fitting Gaussian mixture models to segment and remove observations with low complexity and SNR values (i.e. the tail shown in figure \ref{fig:complexity_snr_results}) and segment the remaining complexity and SNR values to identify the most interesting observations (i.e. simple vs complex). Results are shown in table \ref{tab:confusionmatrixsnrGMM}. 

In assessing the effectiveness of the approach, the most important consideration is the reduction of the type II error rate, measuring the effectiveness of the approach at identifying as many of the interesting observations as possible. Minimising the type I error rate is of importance in providing a significant reduction in the volume of non-interesting data flagged for further investigation. Given a significant reduction in the total data volume and a very low type II error rate, the contamination of the segregated sample as measured by precision was not deemed to be of primary concern. For this reason the Informedness measure, as described in Appendix A, was chosen to assess performance. The Informedness measure incorporates both Type I errors (False Positives) and Type II errors (False Negatives) and describes the improved performance of the approach with respect to chance, costing true positives and false positives in a way analogous to how a bookmaker fairly prices the odds \citep{powers11}. A detailed description of these metrics are provided in Appendix A.

Based on the results shown in table \ref{tab:confusionmatrixsnrGMM} the combined sample (CDFS and ELAIS) produces a recall of $0.88$ (88\% true positives) and a false positive rate of 6\%. This represents an informedness of $0.82$. The CDFS sample provides a recall of $0.86$ and informedness of $0.81$ while the ELAIS sample provides a recall of $0.90$ and informedness of $0.84$. These results show that approximately 90\% of expertly classified complex sources were contained within the largest 10\% of complexity values. 

Results show an improvement in performance when incorporating the SNR, with an improved recall within the CDFS sample and a reduced false positive rate within the ELAIS sample as shown in table \ref{tab:summaryresults}. By reducing the likelihood thresholds used for classification the recall can be further improved at the expense of the false positive rate.

To demonstrate the importance of applying an appropriate smoothing function, an experiment was run to partition the data without smoothing. The images were classified by selecting approximately the top 10\% of observations based on apparent complexity score, as was also done using the smoothed images. The drastic reduction in performance is shown by comparing the results in table \ref{tab:confusionmatrixcomponly} and \ref{tab:confusionmatrixnosmooth}, and comparing the summary measures for the ranked complexity, with and without smoothing, as shown in table \ref{tab:summaryresults}. These results demonstrate the importance of the smoothing function when partitioning interesting observations. Results suggest that the learned smoothing function is successfully able to isolate comprehensible content associated with the meaningful structural information used by astronomers to manually classify the radio sources. 

The results shown in table \ref{tab:summaryresults} also demonstrate that the smoothing function window size learned from the smaller CDFS DR1 sample is able to generalise well when applied to the larger CDFS and ELAIS DR3 samples containing a larger collection of complex morphologies. 

\begin{table}
\begin{tabular}{ l  c  c }
\hline \hline
CDFS (n=2861)& Complex obs & Simple obs\\
\hline 
Prediction: Complex& 83 & 140\\
Prediction: Simple& 14 & 2624\\ 
\hline \hline
ELAIS (n=1964)& Complex obs & Simple obs\\
\hline 
Prediction: Complex& 65 & 126\\
Prediction: Simple& 7 & 1766\\
\hline
\end{tabular}
\caption{\label{tab:confusionmatrixsnrGMM}Confusion matrix (complexity \& SNR GMM partition)}
\end{table}

\begin{table}
\begin{tabular}{ l c c }
\hline \hline
CDFS (n=2861)& Complex obs & Simple obs\\
\hline 
Prediction: Complex& 4 & 321\\
Prediction: Simple& 93 & 2443\\  
\hline \hline
ELAIS (n=1964)& Complex obs & Simple obs\\
\hline 
Prediction: Complex& 2 & 230\\
Prediction: Simple& 70 & 1662\\  
\hline 
\end{tabular}
\caption{Confusion matrix (P90 complexity cut without smoothing)}
\label{tab:confusionmatrixnosmooth}
\end{table}

\begin{table*}
\centering
\begin{tabular}{l c c c}
\hline \hline
\multicolumn{4}{c}{CDFS (n=2861)} \\
\hline 
Method & Recall & False Positive Rate & Informedness \\
\hline 
Percentile cut: complexity without smoothing & 0.04 & 0.12 & -0.07 \\
Percentile cut: coarse-grained complexity & 0.84 & 0.09 & 0.75 \\
GMM: coarse-grained complexity & 0.75 & 0.05 & 0.70 \\
GMM: coarse-grained complexity and SNR & 0.86 & 0.05 & 0.81 \\
\hline \hline 
 \multicolumn{4}{c}{ELAIS (n=1964)} \\
\hline 
Method & Recall & False Positive Rate & Informedness \\
\hline 
Percentile cut: complexity without smoothing & 0.03 & 0.12 & -0.09 \\
Percentile cut: coarse-grained complexity & 0.90 & 0.09 & 0.81 \\
GMM: coarse-grained complexity & 0.90 & 0.08 & 0.82 \\
GMM: coarse-grained complexity and SNR & 0.90 & 0.07 & 0.84 \\
\hline 
\end{tabular}
\caption{\label{tab:summaryresults} Summary of experimental results, including: Recall, False Positive Rate and informedness measures.}
\end{table*}

\subsection{Validation using synthetic data}

\begin{figure}[h]
\includegraphics[width=8cm]{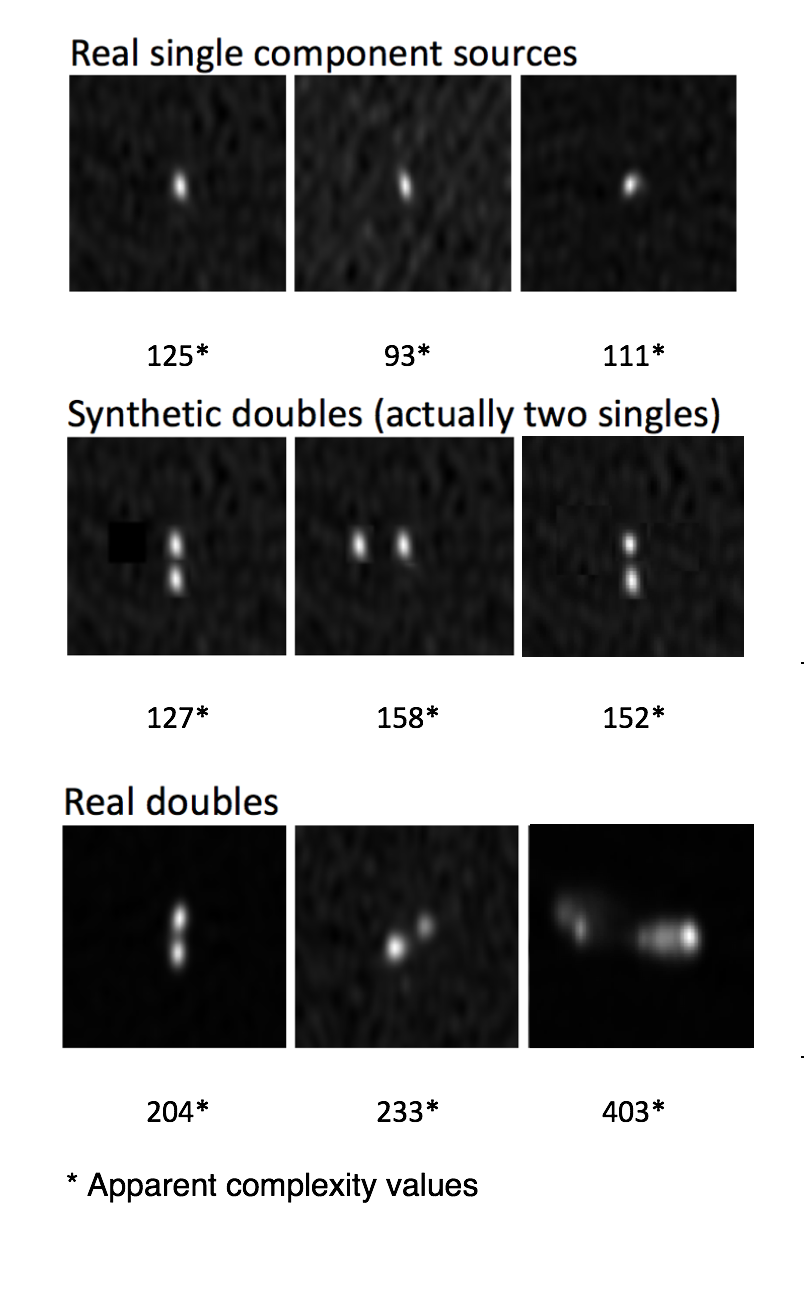}
\caption{\label{fig:testcase}The apparent complexity (in bytes) calculated for images containing only simple radio sources, synthetic sources combining two simple sources and real complex radio sources. These results show that the apparent complexity values can be used to distinguish between the true complex sources and simple sources in close proximity.}
\end{figure}

Synthetic radio images were constructed by placing simple radio sources in close proximity, comparable to the distances between components associated with real complex sources with two components (i.e. real doubles). In accordance with the unique decompression property the distinct components of true complex sources should produce larger apparent complexity values. The apparent complexity was calculated for images containing only simple radio sources, synthetic sources combining two simple sources and real complex radio sources as shown in figure \ref{fig:testcase}. The apparent complexity values distinguish between the true complex sources and synthetic sources consisting of two simple sources in close proximity, with true complex sources producing larger complexity values.

\subsection{Method Comparisons}

The proposed approach presents two key attributes. Firstly, apparent complexity is fast and computationally efficient to compute. The measure requires only applying a smoothing function and compression algorithm, both of which can be implemented to scale at worst case linear time complexity, making the measure appropriate for analysing large datasets. Secondly, the approach generalises well when using only a very small training set to calibrate the smoothing function. Combined, these attributes make the approach a good candidate for detecting the unexpected in large data produced from new astronomical surveys. These surveys will produce data where the volume, and likely scope, of observations will greatly exceed any existing data available for training classifiers.

However, the question remains as to whether this approach performs as reliably as traditional binary classification approaches applied to pixel data. We tested our method,  comparing the proposed approach to more traditional classification methods, namely, logistic regression and support vector machines (SVM). 

Feature preparation for logistic regression and SVM training and testing involved flattening the images and sorting the resulting one-dimensional arrays by pixel intensity value. The ranked or distributional representation of pixel values are more suitable for classification, as they are invariant to rigid motions of the plane. Logistic regression was implemented with the loss function optimised using gradient descent\footnote{Here we implement gradient descent using the {\tt Tensorflow} package \citep{tensorflow2015-whitepaper} having achieved superior performance to the implementation of the Library for Large Linear Classification (LIBLINEAR) using {\tt scikit-learn} \citep{scikit-learn}}. The SVM was implemented using a 2 degree polynomial kernel specified with a coefficient 1. 

Training and testing on ATLAS DR1 data was performed using cross-validation, splitting the DR1 set in two (separate training and testing sets) each time,  with each set of 354 images containing 17 complex images. Thus the training data was comparable to the range of complex observations in the ATLAS DR1 data that was used for testing. These experiments employed the same data used in other studies \citep{park2018, alger2018}, to allow comparison to additional approaches including approaches that incorporate feature learning such as Convolution Neural Networks. The approaches tested here achieved comparable performance to the approaches tested by \citet{park2018} in the binary classification of simple and complex morphology. 

Table \ref{tab:comparisonresults} shows results from experiments using logistic regression and SVM models implemented using the ranked pixel data. In terms of the informedness and recall metrics, these binary classification approaches achieve results comparable to the method using the coarse-grained complexity measure. 

Having trained on ATLAS DR1 (n=708), we then tested using ATLAS DR3 (n=4825). Results are also shown in table \ref{tab:comparisonresults}. We found that logistic regression applied to the ranked pixel data provided an informedness of 0.64 (across both CDFS and ELAIS samples) and the SVM applied to the ranked pixel data provided an informedness of 0.48, both of which were considerably smaller than the informedness of 0.82 produced when using the apparent complexity measure. Thus the complexity measure was better able to generalise when using a smaller training set. This shows that the smoothing scale can be effectively calibrated using only a small number of complex images to distinguish between simple images and a broad range of complex images. In contrast, the classification methods applied to the pixel data were unable to replicate this performance, being `under-trained' for such an expanded data set.

These tests demonstrate that the proposed approach provides comparable performance to traditional classification methods applied to pixel data when the training data is representative of the range of complex observations in the testing data. They also demonstrate however that the use of the proposed complexity measure provides superior performance to the traditional classification methods when the training dataset is much smaller than the testing dataset and the training data contains only a small sample of the range of complex observations found in the testing data. This later scenario is representative of the challenge that comes with larger astronomical surveys.

\begin{table*}
\centering
\begin{tabular}{l c c c}
\hline \hline
 \multicolumn{4}{c}{DR1 CDFS (n=708 cross val)} \\
\hline 
Method & Recall & False Positive Rate & Informedness \\
\hline 
Logistic regression & 1.00 & 0.19 & 0.81 \\
SVM & 0.88 & 0.02 & 0.86 \\
GMM: coarse-grained complexity & 0.85 & 0.06 & 0.79 
\\
\hline \hline 
 \multicolumn{4}{c}{ DR3 CDFS (n=2861)} \\
\hline 
Method & Recall & False Positive Rate & Informedness \\
\hline 
Logistic regression & 0.99 & 0.36 & 0.63 \\
SVM & 0.61 & 0.03 & 0.58 \\
GMM: coarse-grained complexity & 0.75 & 0.05 & 0.70 \\
GMM: coarse-grained complexity and SNR & 0.86 & 0.05 & 0.81 \\
\hline \hline 
 \multicolumn{4}{c}{ DR3 ELAIS (n=1964)} \\
\hline 
Method & Recall & False Positive Rate & Informedness \\
\hline 
Logistic regression & 0.94 & 0.30 & 0.65 \\
SVM & 0.36 & 0.02 & 0.34 \\
GMM: coarse-grained complexity & 0.90 & 0.08 & 0.82 \\
GMM: coarse-grained complexity and SNR & 0.90 & 0.07 & 0.84 \\
\hline 
\end{tabular}
\caption{\label{tab:comparisonresults} Classification performance of GMM using calibrated coarse-grained complexity compared to alternative classification approaches.}
\end{table*}

\section{Conclusions}
\label{sec:conclusions}

We introduce the apparent complexity as a measure to identify interesting or complex observations in large astronomical datasets. The apparent complexity is defined as the entropy of an image after applying a smoothing function, and is approximated by the file size of a corresponding image after smoothing and compression. The measure can be computed at worst case linear time complexity, making it well suited for analysing very large datasets to identify smaller and relevant samples for additional analysis requiring more computationally intensive methods or manual inspection.

Using imaging data from the ATLAS radio continuum survey, we have shown that apparent complexity, combined with the signal-to-noise ratio, can be used to partition images of galaxies into those with simple and complex morphologies. Partitioning the images by fitting Gaussian mixture models using the expectation maximisation algorithm, we are able to implement an automated process to identify complex sources with a recall of 0.88 and informedness of 0.82 across both CDFS and ELAIS samples. Results also show that approximately 90\% of expertly classified complex sources were contained within the largest 10\% of complexity values. 

We demonstrate that the association between apparent complexity and the morphological complexity of observations generalises well, positioning the measure as a candidate machine learning feature for identifying complex observations with unanticipated morphological features (i.e. unknown unknowns) in the presence of noise. We have shown that the apparent complexity with a smoothing function window size learned from the smaller ATLAS DR1 sample (n=720) is able to generalise well when applied to the much larger ATLAS DR3 sample (n=4825) containing a larger collection of complex morphologies. The use of apparent complexity to segment images and identify interesting observations should generalise well across other samples, including different types of observational data outside of radio frequencies. Further testing with large radio samples and new types of data will be needed to test this hypothesis.

Experiments were conducted comparing the proposed approach using the apparent complexity measure against more traditional classification methods, namely logistic regression and support vector machines (SVM), applied to pixel data. These experiments demonstrated that the use of the coarse-grained complexity measure provides comparatively superior classification performance when the training data contains only a small sample of the complex observations found in the testing data. Training on ATLAS DR1 (n=708) and testing on ATLAS DR3 (n=4825), logistic regression applied to pixel data provided an informedness of 0.64 and the SVM applied to pixel data provided an informedness of 0.48, both being smaller than the informedness of 0.82 produced when using the apparent complexity measure.

We envisage a potential application for the apparent complexity measure at the early stages of the analysis pipeline for radio images. The early identification of complex radio sources allows these observations to be segmented for additional analysis. This is likely to include cross matching with observations at other frequencies to determine if the host galaxy is coincident with any of the radio components. It may also serve in the early identification of new and unexpected observations that may arise as part of next generation surveys, for example the Square Kilometre Array \citep{SKA2015Jarvis,ska_unknown_unknowns}. 

While a large number of the most interesting observations are likely to be more complex than less interesting observations, we do acknowledge that there will be interesting outliers in scientific datasets that are relatively simple. Further work to understand the intersection of the sets of outlying and complex observations will clarify the potential limitations and best applications of a complexity based approach. 

\section*{Acknowledgements}

The authors would like to thank Geoffrey McLachlan, Laurence Park, Evan Crawford and Michael Drinkwater for some helpful discussions and early contributions.

\appendix
\section{Evaluation measures}
\label{sec:informedness}

For a binary classification problem a 2x2 contingency table can be constructed to represent counts of False Positives (FP), True Positives (TP), False Negatives (FN) and True Negatives (TN) as depicted by table \ref{tab:confusionmatrix}. Such a table is referred to as a Confusion Matrix and depicts both the counts of Type I errors (False Positives) and Type II errors (False Negatives). 

\begin{table} [h]
\centering
\begin{tabular}{c c c c} \hline \hline
 & \textbf{Class +} &  \textbf{Class -} & \textbf{Total} \\ \cline{1-4} 
\textbf{Prediction +} & TP & FP (Type I error) & Predicted Positives (PP) \\ \cline{1-4}
\textbf{Prediction -} & FN (Type II error) & TN  &  Predicted Negatives (PN) \\ \cline{1-4}
\textbf{Total} & Real Positives (RP) & Real Negatives (RN)  &  \\ \cline{1-4}
\end{tabular}
\caption{\label{tab:confusionmatrix}Confusion Matrix for binary classification problem}
\end{table}

Quantities can be derived using the information contained within a binary Confusion Matrix to measure the performance of a classifier. Two such quantities are precision
\begin{equation}
{\rm Precision} = \frac{TP}{TP+FP}\,,
\end{equation}
and  recall
\begin{equation}
{\rm Recall} = \frac{TP}{TP+FN}\,.
\end{equation}
Precision determines the number of correct positive classifications as a fraction of positive classifications, while recall determines the number of correct positive classifications as a fraction of the total number of actual positives (so the fraction of positive objects that have been missed would be $1-{\rm recall}$, in the binary classification case). 

An alternative framework for measuring performance involves the use of Receiver Operating Characteristic (ROC) curves. The use of ROC curves to construct a comparative framework has been adopted in the machine learning literature \citep{Flach05}. These approaches account for chance level performance and can also be use to account for the cost weightings of negative and positive cases. ROC analysis examines the false positive rate (FP/RN) versus the true positive rate (TP/RP) and presents equivalent information to ratios calculated in the vertical direction of the Confusion Matrix presented in table \ref{tab:confusionmatrix}. 

The maximum distance of the receiver operating characteristic (ROC) curve from the 45 degree chance line is equivalent to Youden's J statistic \citep{Youden} and the Informedness measure \citep{powers11}. The Informedness measure is equivalent to the subtraction of the false positive rate (FPR) from the true positive rate (TPR) as follows:

	\begin{equation} \textbf{Informedness} = \frac{TP}{TP+FN} - \frac{FP}{TN+FP} = TPR - FPR
\end{equation}

This measure is also equivalent to a chance adjusted version of Recall:

\begin{equation} \textbf{Informedness} = {\rm recall} - FPR
\end{equation}

\citet{powers11} shows that Informedness is an unbiased estimator of above chance performance. The measure incorporates both Type I errors (False Positives) and Type II errors (False Negatives) and describes the improved performance of the measured classifier with respect to chance, costing true positives and false positives in a way analogous to how a bookmaker fairly prices the odds \citep{powers11}. For this reason the measure is also referred to as Bookmaker Informedness. The Informedness measure is defined on a (-1,1) interval and gives equal weighting to the true positive and false positive rate.

Informedness appears appropriate for evaluating the effectiveness of alternative approaches at detecting and classifying interesting observations in large astronomical data. The Informedness measure relates to the following objectives of classification:

\begin{enumerate}
	\item 	\textbf{Maximise true positive rate} (i.e. minimise the type II error rate) - providing assurance that actual interesting observations are available for analysis.
	\item 	\textbf{Minimise false positive rate} (i.e. minimise the type I error rate) - to minimise the data burden and place minimal unnecessary burden on data transmission and storage infrastructure.
\end{enumerate} 

Removing false positives reduces storage and data handling requirements and the associated costs. Retaining a smaller subset of observations that are likely to be interesting allows these to be directed to low latency storage options where they can be easily retrieved. 

Due to the likely small number of actual interesting observations compared to normal observations the metric is likely to be more sensitive to small changes in the true positive count and less sensitive to small changes in the false positive count.

\section{Sensitivity analysis: Pixel intensity threshold}
\label{sec:intensity threshold}

We consider a range of possible values for the threshold pixel intensity, that removes random information from the image, and  preserves the non-random or `true' image information particular to the object.
Figure \ref{fig:thresholds} shows the differences in average apparent complexity between simple and complex ATLAS DR1 images at different percentile based pixel intensity thresholds, namely, the P40 and P50, P60, P70, P80 and P90, as a function of smoothing scale. At thresholds lower below P60 the images become noise dominated and effective separation between simple and complex images is no longer possible through smoothing. The simple images produce larger complexity values attributable to noise. At all possible threshold values, the effective separation between simple and complex images disappears as the smoothing scale greatly exceeds the size of source features. These results support the appropriateness of a large threshold, such as the P90, which provides clear separation between simple and complex sources and larger average apparent complexity for complex sources in the presence of noise. This parameter has not been finely tuned however, and alternative threshold values may provide superior performance. Furthermore, this value was held fixed during calibration and testing so that the smoothing resolution was the only free parameter. 
\begin{table}[!htb] 
\centering
\begin{tabular}{c c}
\hline
\includegraphics[width=75mm]{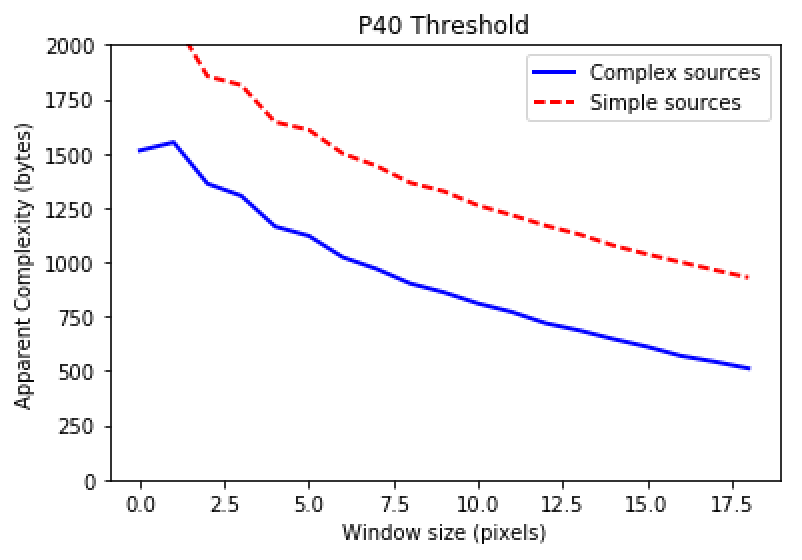} &
\includegraphics[width=75mm]{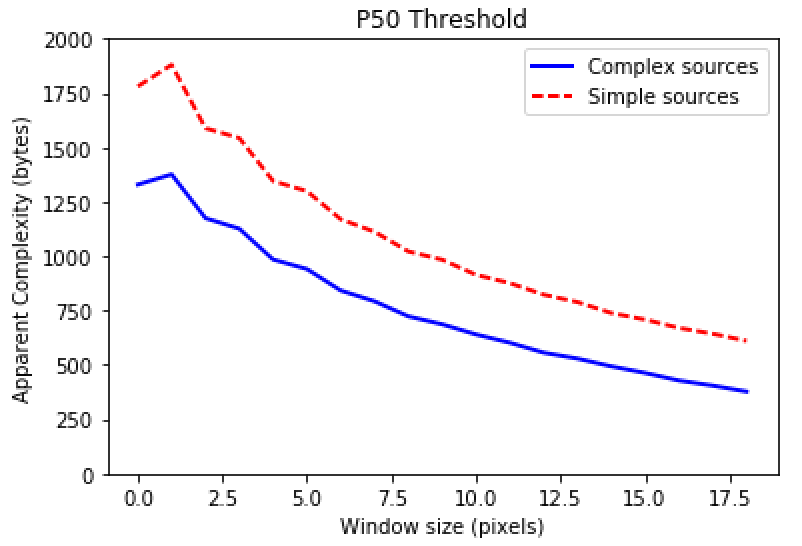}
\\ 
\hline
\includegraphics[width=75mm]{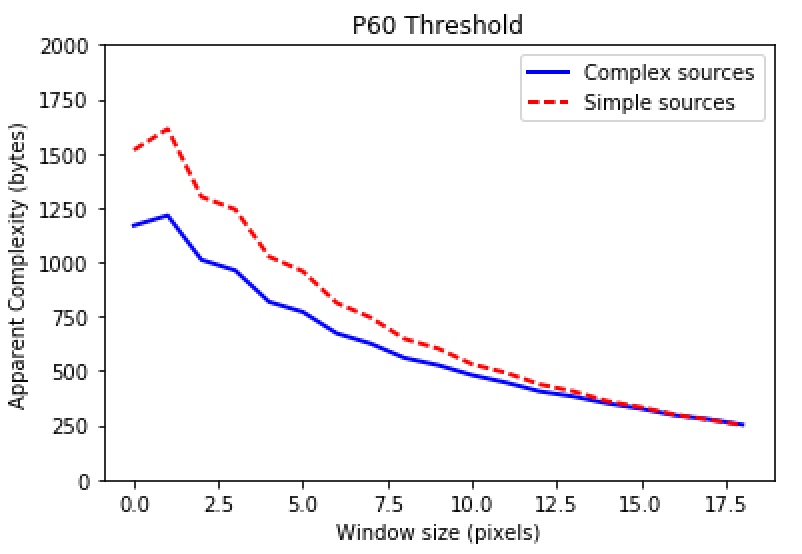} &
\includegraphics[width=75mm]{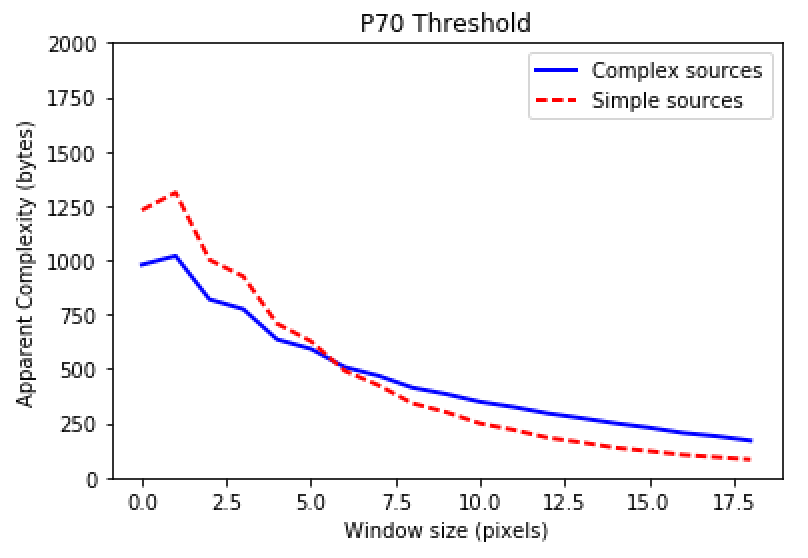}
\\
\hline
\includegraphics[width=75mm]{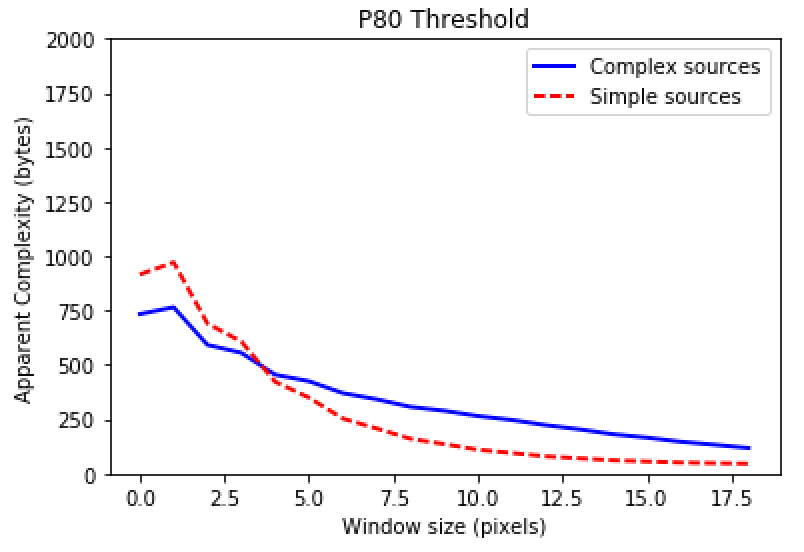} & 
\includegraphics[width=75mm]{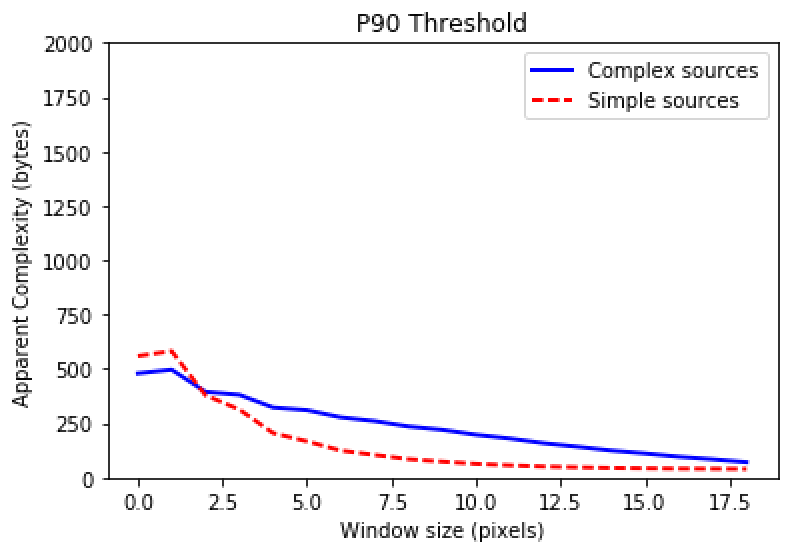}
\\
\hline
\end{tabular}
\caption{\label{fig:thresholds} Plots showing the differences in average apparent complexity between simple and complex ATLAS DR1 images as a function of smoothing scale at different percentile based pixel intensity thresholds.}
\end{table}

\pagebreak
\bibliographystyle{aasjournal}
\bibliography{refs}
\end{document}